\documentclass{jpconf}

\usepackage{graphicx}
\usepackage{amsmath}
\usepackage{amsfonts}
\usepackage{mathrsfs}
\usepackage{cite}
\usepackage{epsfig}
\usepackage{picture,epic}

\newcommand{\beqs}{\begin{equation*}}
\newcommand{\beq}{\begin{equation}}

\newcommand{\eeqs}{\end{equation*}}
\newcommand{\eeq}{\end{equation}}

\newcommand{\beqas}{\begin{eqnarray*}}
\newcommand{\beqa}{\begin{eqnarray}}

\newcommand{\eeqas}{\end{eqnarray*}}
\newcommand{\eeqa}{\end{eqnarray}}

\newcommand{\eq}[2]{\begin{equation} #1 \label{#2} \end{equation}}

\newcommand{\eps}{\varepsilon}
\newcommand{\al}{\alpha}
\newcommand{\be}{\beta}
\newcommand{\ga}{\gamma}
\newcommand{\de}{\delta}

\newcommand{\la}{\lambda}
\newcommand{\si}{\sigma}

\newcommand{\Ga}{\Gamma}
\newcommand{\De}{\Delta}

\newcommand{\blist}{\begin{itemize}}

\newcommand{\elist}{\end{itemize}}

\providecommand{\href}[2]{#2}

\DeclareFontFamily{OT1}{rsfs}{}
\DeclareFontShape{OT1}{rsfs}{m}{n}{ <-7> rsfs5 <7-10> rsfs7 <10->rsfs10}{} 
\DeclareMathAlphabet{\mycal}{OT1}{rsfs}{m}{n}

\DeclareMathOperator{\extdm}{d}
\newcommand{\extd}{\extdm \!}

\newcommand\MM{\mathcal{M}}
\newcommand\dM{\partial \MM}

\newcommand{\volint}{\!\extd^3x}

\begin{document}

\title{Gravity duals for logarithmic conformal field theories}

\author{Daniel Grumiller and Niklas Johansson}

\address{Institute for Theoretical Physics, Vienna University of Technology\\ Wiedner Hauptstrasse 8-10/136, A-1040 Vienna, Austria}

\ead{grumil@hep.itp.tuwien.ac.at, niklasj@hep.itp.tuwien.ac.at}

\begin{abstract}
Logarithmic conformal field theories with vanishing central charge describe systems with quenched disorder, percolation or dilute self-avoiding polymers. In these theories the energy momentum tensor acquires a logarithmic partner. In this talk we address the construction of possible gravity duals for these logarithmic conformal field theories and present two viable candidates for such duals, namely theories of massive gravity in three dimensions at a chiral point. 
\end{abstract}

\section*{Outline}

This talk is organized as follows. In section \ref{sec:intro} we recall salient features of 2-dimensional conformal field theories. In section \ref{sec:1} we review a specific class of logarithmic conformal field theories where the energy momentum tensor acquires a logarithmic partner. In section \ref{sec:2} we present a wish-list for gravity duals to logarithmic conformal field theories. In section \ref{sec:3} we discuss two examples of massive gravity theories that comply with all the items on that list. In section \ref{sec:4} we address possible applications of an Anti-deSitter/logarithmic conformal field theory correspondence in condensed matter physics.

\section{Conformal field theory distillate}\label{sec:intro}

Conformal field theories (CFTs) are quantum field theories that exhibit invariance under angle preserving transformations: translations, rotations, boosts, dilatations and special conformal transformations. In two dimensions the conformal algebra is infinite dimensional, and thus two-dimensional CFTs exhibit a particularly rich structure. They arise in various contexts in physics, including string theory, statistical mechanics and condensed matter physics, see e.g.~\cite{diFrancesco}. 

The main observables in any field theory are correlation functions between gauge invariant operators. There exist powerful tools to calculate these correlators in a CFT. The operator content of various CFTs may differ, but all CFTs contain at least an energy momentum tensor $T_{\mu\nu}$. Conformal invariance requires the energy momentum tensor to be traceless, $T^\mu_\mu=0$, in addition to its conservation, $\partial_\mu T^{\mu\nu}=0$. In lightcone gauge for the Minkowski metric, $\extd s^2=2\extd z\extd\bar z$, these equations take a particularly simple form: $T_{z\bar z}=0$, $T_{zz}=T_{zz}(z):={\cal O}^L(z)$ and $T_{\bar z\bar z}=T_{\bar z\bar z}(\bar z):={\cal O}^R(\bar z)$. Conformal Ward identities determine essentially uniquely the form of 2- and 3-point correlators between the flux components ${\cal O}^{L/R}$ of the energy momentum tensor:
\begin{subequations}
\label{eq:LCFT1}
\begin{align}
& \langle{\cal O}^R(\bar z)\,{\cal O}^R(0)\rangle = \frac{c_R}{2\bar z^4} \\
& \langle{\cal O}^L(z)\,{\cal O}^L(0)\rangle = \frac{c_L}{2 z^4} \\
& \langle{\cal O}^L(z)\,{\cal O}^R(0)\rangle = 0 \\
& \langle{\cal O}^R(\bar z)\,{\cal O}^R(\bar z')\,{\cal O}^R(0)\rangle = \frac{c_R}{\bar z^2\bar z^{\prime\,2}(\bar z-\bar z')^2} \\
& \langle{\cal O}^L(z)\,{\cal O}^L(z')\,{\cal O}^L(0)\rangle = \frac{c_L}{z^2z^{\prime\,2}(z-z')^2} \\
& \langle{\cal O}^L(z)\,{\cal O}^R(\bar z')\,{\cal O}^R(0)\rangle = 0 \\
& \langle{\cal O}^L(z)\,{\cal O}^L(z')\,{\cal O}^R(0)\rangle = 0 
\end{align}
\end{subequations}
The real numbers $c_L$, $c_R$ are the left and right central charges, which determine key properties of the CFT. We have omitted terms that are less divergent in the near coincidence limit $z,\bar z\to 0$ as well as contact terms, i.e., contributions that are localized ($\de$-functions and derivatives thereof).

If someone provides us with a traceless energy momentum tensor and gives us a prescription how to calculate correlators,\footnote{This is exactly what the AdS/CFT correspondence does: given a gravity dual we can calculate the energy momentum tensor and correlators.} but does not reveal whether the underlying field theory is a CFT, then we can perform the following check. We calculate all 2- and 3-point correlators of the energy momentum tensor with itself, and if at least one of the correlators does not match precisely with the corresponding correlator in \eqref{eq:LCFT1} then we know that the field theory in question cannot be a CFT. On the other hand, if all the correlators match with corresponding ones in \eqref{eq:LCFT1} we have non-trivial evidence that the field theory in question might be a CFT. Let us keep this stringent check in mind for later purposes, but switch gears now and consider a specific class of CFTs, namely logarithmic CFTs (LCFTs).

%This talk is organized as follows. In section \ref{sec:1} we review a specific class of LCFTs where the energy momentum tensor acquires a logarithmic partner. In section \ref{sec:2} we present a wish-list for gravity duals to LCFTs. In section \ref{sec:3} we discuss two examples of massive gravity theories that comply with all the items on that list. In section \ref{sec:4} we address possible applications of an AdS/LCFT correspondence in condensed matter physics.

\section{Logarithmic CFTs with an energetic partner}\label{sec:1}

LCFTs were introduced in physics by Gurarie \cite{Gurarie:1993xq}. We focus now on some properties of LCFTs and postpone a physics discussion until the end of the talk, see \cite{Flohr:2001zs,Gaberdiel:2001tr} for reviews. There are two conceptually different, but mathematically equivalent, ways to define LCFTs. In both versions there exists at least one operator that acquires a logarithmic partner, which we denote by ${\cal O}^{\rm log}$. We focus in this talk exclusively on theories where one (or both) of the energy momentum tensor flux components is the operator acquiring such a partner, for instance ${\cal O}^L$. We discuss now briefly both ways of defining LCFTs.

According to the first definition ``acquiring a logarithmic partner'' means that the Hamiltonian $H$ cannot be diagonalized. For example
\eq{ 
H \left(\begin{array}{c} {\cal O}^{\rm log} \\ {\cal O}^L
\end{array}\right) = \left(\begin{array}{c@{\quad}c}
2 & 1 \\
0 & 2
\end{array}\right) \left(\begin{array}{c} {\cal O}^{\rm log} \\ {\cal O}^L \end{array}\right)
}{eq:cg79} 
The angular momentum operator $J$ may or may not be diagonalizable. We consider only theories where $J$ is diagonalizable:
\eq{ 
J \left(\begin{array}{c} {\cal O}^{\rm log} \\ {\cal O}^L
\end{array}\right) = \left(\begin{array}{c@{\quad}c}
2 & 0 \\
0 & 2
\end{array}\right) \left(\begin{array}{c} {\cal O}^{\rm log} \\ {\cal O}^L \end{array}\right)
}{eq:cg1} 
The eigenvalues $2$ arise because the energy momentum tensor and its logarithmic partner both correspond to spin-2 excitations.

The second definition makes it more transparent why these CFTs  are called ``logarithmic'' in the first place. % if they contain operators with degenerate scaling dimensions having a Jordan block structure like in \eqref{eq:cg79}. 
Suppose that in addition to ${\cal O}^{L/R}$ we have an operator ${\cal O}^M$ with conformal weights $h=2+\eps$, $\bar h=\eps$, meaning that its 2-point correlator with itself is given by
\eq{
\langle {\cal O}^M(z,\bar z)\,{\cal O}^M(0,0)\rangle = \frac{\hat B}{z^{4+2\eps}\bar z^{2\eps}}
}{eq:LCFT2}
The correlator of ${\cal O}^M$ with ${\cal O}^L$ vanishes since the latter has conformal weights $h=2$, $\bar h=0$, and operators whose conformal weights do not match lead to vanishing correlators. Suppose now that we send the central charge $c_L$ and the parameter $\eps$ to zero, and simultaneously send $\hat B$ to infinity, such that the following limits exist:
\eq{
b_L:=\lim_{c_L\to 0} -\frac{c_L}{\eps}\neq 0 \qquad B:=\lim_{c_L\to 0} \big(\hat B + \frac{2}{c_L}\big)
}{eq:LCFT3}
Then we can define a new operator ${\cal O}^{\rm log}$ that linearly combines ${\cal O}^{L/M}$.
\eq{
{\cal O}^{\rm log} = b_L\,\frac{{\cal O}^L}{c_L}+\frac{b_L}{2}\,{\cal O}^M
}{eq:LCFT4}
Taking the limit $c_L\to 0$ leads to the following 2-point correlators:
\begin{subequations}
\label{eq:2point}
\begin{align}
& \langle{\cal O}^L(z){\cal O}^L(0,0)\rangle = 0 \label{eq:LL} \\
& \langle{\cal O}^L(z){\cal O}^{\rm log}(0,0)\rangle = \frac{b_L}{2z^4} \label{eq:Llog} \\
& \langle{\cal O}^{\rm log}(z,\bar z){\cal O}^{\rm log}(0,0)\rangle = -\frac{b_L \ln{(m^2_L|z|^2)}}{z^4} \label{eq:loglog}
\end{align}
\end{subequations}
These 2-point correlators exhibit several remarkable features. The flux component ${\cal O}^L$ of the energy momentum tensor becomes a zero norm state \eqref{eq:LL}. Nevertheless, the theory does not become chiral, because the left-moving sector is not trivial: ${\cal O}^L$ has a non-vanishing correlator \eqref{eq:Llog} with its logarithmic partner ${\cal O}^{\rm log}$.  The 2-point correlator \eqref{eq:loglog} between two logarithmic operators ${\cal O}^{\rm log}$ makes it clear why such CFTs have the attribute ``logarithmic''. The constant $b_L$, sometimes called ``new anomaly'', defines crucial properties of the LCFT, much like the central charges do in ordinary CFTs. The mass scale $m_L$ appearing in the last correlator above has no significance, and is determined by the value of $B$ in \eqref{eq:LCFT3}. It can be changed to any finite value by the redefinition ${\cal O}^{\rm log}\to{\cal O}^{\rm log}+\ga\,{\cal O}^L$ with some finite $\ga$. We set $m_L=1$ for convenience.

Conformal Ward identities determine again essentially uniquely the form of 2- and 3-point correlators in a LCFT. For the specific case where the energy momentum tensor acquires a logarithmic partner the 3-point correlators were calculated in \cite{Kogan:2001ku}. The non-vanishing ones are given by
\begin{subequations}
\label{eq:3point}
\begin{align}
& \langle{\cal O}^L(z,\bar z){\cal O}^L(z',\bar z'){\cal O}^{\rm log}(0,0)\rangle = \frac{b_L}{z^2{z'}^2(z-z')^2} \\
& \langle{\cal O}^L(z,\bar z){\cal O}^{\rm log}(z',\bar z'){\cal O}^{\rm log}(0,0)\rangle = -\frac{2b_L\ln{|{z'}|^2}+\frac{b_L}{2}}{z^2{z'}^2(z-z')^2} \\
& \langle{\cal O}^{\rm log}(z,\bar z){\cal O}^{\rm log}(z',\bar z'){\cal O}^{\rm log}(0,0)\rangle = \frac{{\rm lengthy}}{z^2{z'}^2(z-z')^2}
\end{align}
\end{subequations}
If also ${\cal O}^R$ acquires a logarithmic partner ${\cal O}^{\widetilde{\rm log}}$ then the construction above can be repeated, changing everywhere $L\to R$, $z\to \bar z$ etc. In that case we have a LCFT with $c_L=c_R=0$ and $b_L,b_R\neq 0$. Alternatively, it may happen that only ${\cal O}^L$ has a logarithmic partner ${\cal O}^{\rm log}$. In that case we have a LCFT with $c_L=b_R=0$ and $b_L,c_R\neq 0$. This concludes our brief excursion into the realm of LCFTs.

Given that LCFTs are interesting in physics (see section \ref{sec:4}) and that a powerful way to describe strongly coupled CFTs is to exploit the AdS/CFT correspondence \cite{Aharony:1999ti} it is natural to inquire whether there are any gravity duals to LCFTs.

\section{Wish-list for gravity duals to LCFTs}\label{sec:2}

In this section we establish necessary properties required for gravity duals to LCFTs. We formulate them as a wish-list and explain afterwards each item on this list.
\begin{enumerate}
\item We wish for a 3-dimensional action $S$ that depends on the metric $g_{\mu\nu}$ and possibly on further fields that we summarily denote by $\phi$.
\item We wish for the existence of AdS$_3$ vacua with finite AdS radius $\ell$.
\item We wish for a finite, conserved and traceless Brown--York stress tensor, given by the first variation of the full on-shell action (including boundary terms) with respect to the metric.
\item We wish that the 2- and 3-point correlators of the Brown--York stress tensor with itself are given by \eqref{eq:LCFT1}.
\item We wish for central charges (a la Brown--Henneaux \cite{Brown:1986nw}) that can be tuned to zero, without requiring a singular limit of the AdS radius or of Newton's constant. For concreteness we assume $c_L=0$ (in addition $c_R$ may also vanish, but it need not).
\item We wish for a logarithmic partner to the Brown--York stress tensor, so that we obtain a Jordan-block structure like in \eqref{eq:cg79} and \eqref{eq:cg1}.
%\item We wish that $c_L=0$ (in addition $c_R$ may also vanish, but it need not).
\item We wish  that the 2- and non-vanishing 3-point correlators of the Brown--York stress tensor with its logarithmic partner are given by \eqref{eq:2point} and \eqref{eq:3point} (and the right-handed analog thereof).
\end{enumerate}
We explain now why each of these items is necessary. (i) is required since the AdS/CFT correspondence relates a gravity theory in $d+1$ dimensions to a CFT in $d$ dimensions, and we chose $d=2$ on the CFT side. (ii) is required since we are not merely looking for a gauge/gravity duality, but really for an AdS/CFT correspondence, which requires the existence of AdS solutions on the gravity side. (iii) is required since we desire consistency with the AdS dictionary, which relates the vacuum expectation value of the renormalized energy momentum tensor in the CFT $\langle T_{ij}\rangle$ to the Brown--York stress tensor $T_{ij}^{\rm BY}$:
\eq{
\langle T_{ij}\rangle = T_{ij}^{\rm BY} = \frac{2}{\sqrt{-g}}\,\frac{\de S}{\de g^{ij}}\Big|_{\textrm{\tiny EOM}}
}{eq:LCFT13}
The right hand side of this equation contains the first variation of the full on-shell action with respect to the metric, which by definition yields the Brown--York stress tensor. (iv) is required since the 2- and 3-point correlators of a CFT are fixed by conformal Ward identities to take the form \eqref{eq:LCFT1}. (v) is required because of the construction presented in section \ref{sec:1}, where a LCFT emerges from taking an appropriate limit of vanishing central charge, so we need to be able to tune the central charge without generating parametric singularities.  Actually, there are two cases: either left and right central charge vanish and both energy momentum tensor flux components acquire a logarithmic partner, or only one of them acquires a logarithmic partner, which for sake of specificity we always choose to be left. (vi) is required, since we consider exclusively LCFTs where the energy momentum tensor acquires a logarithmic partner. (vii) is required  since the 2- and 3-point correlators of a LCFT are fixed by conformal Ward identities to take the form \eqref{eq:2point}, \eqref{eq:3point}. 
If any of the items on the wish-list above is not fulfilled it is impossible that the gravitational theory under consideration is a gravity dual to a LCFT of the type discussed in section \ref{sec:1}.\footnote{Other types of LCFTs exist, e.g.~with non-vanishing central charge or with logarithmic partners to operators other than the energy momentum tensor. The gravity duals for such LCFTs need not comply with all the items on our wish list.} % In particular, it is in general not necessary that the gravitational part of the action contains higher-derivative interactions.} 
On the other hand, if all the wishes are granted by a given gravitational theory there are excellent chances that this theory is dual to a LCFT. Until recently no good gravity duals for LCFTs were known \cite{Ghezelbash:1998rj,Myung:1999nd,Kogan:1999bn,Lewis:1999qv,MoghimiAraghi:2004ds}.

Before addressing candidate theories that may comply with all wishes we review briefly how to calculate correlators on the gravity side  \cite{Aharony:1999ti}, since we shall need such calculations for checking several items on the wish-list. The basic identity of the AdS/CFT dictionary is
\eq{
\langle{\cal O}_1(z_1)\,{\cal O}_2(z_2) \dots {\cal O}_n(z_n)\rangle = \frac{\de^{(n)} S}{\de j_1(z_1) \de j_2(z_2) \dots \de j_n(z_n)}\Big|_{j_i=0}
}{eq:LCFT10}
The left hand side is the CFT correlator between $n$ operators ${\cal O}_i$, where ${\cal O}_i$ in our case comprise the left- and right-moving flux components of the energy momentum tensor and their logarithmic partners. The right hand side contains the gravitational action $S$ differentiated with respect to appropriate sources $j_i$ for the corresponding operators. According to the AdS/CFT dictionary ``appropriate sources'' refers to non-normalizable solutions of the linearized equations of motion. We shall be more concrete about the operators, actions, sources and non-normalizable solutions to the linearized equations of motion in the next section. For now we address possible candidate theories of gravity duals to LCFTs.

The simplest candidate, pure 3-dimensional Einstein gravity with a cosmological constant described by the action
\eq{
S_{\textrm{\tiny EH}} = -\frac{1}{8\pi\,G_N}\,\int_{\MM}\!\!\volint\sqrt{-g}\,\Big[R+\frac{2}{\ell^2}\Big] -\frac{1}{4\pi\,G_N}\,\int_{\dM}\!\!\!\!\!\!\extd^2x\sqrt{-\ga}\,\Big[K-\frac{1}{\ell}\Big] 
}{eq:f2}
does not comply with the whole wish list. Only the first four wishes are granted: The 3-dimensional action \eqref{eq:cg20} depends on the metric. The equations of motion are solved by AdS$_3$.
\eq{
\extd s^2_{{\rm AdS}_3} = g^{\textrm{\tiny AdS}_3}_{\mu\nu}\,\extd x^\mu\extd x^\nu = \ell^2\,\big(\extd\rho^2-\frac14\,\cosh^2{\!\!\rho}\, (\extd u+\extd v)^2 +\frac14\,\sinh^2{\!\!\rho}\,(\extd u-\extd v)^2\big)
}{eq:cg20}
The Brown--York stress tensor \eqref{eq:LCFT13} is finite, conserved and traceless. The 2- and 3-point correlators on the gravity side match precisely with \eqref{eq:LCFT1}. However, the central charges are given by \cite{Brown:1986nw}
\eq{
c_L=c_R=\frac{3\ell}{2G_N}
}{eq:LCFT5}
and therefore allow no tuning to $c_L=0$ without taking a singular limit. Moreover, there is no candidate for a logarithmic partner to the Brown--York stress tensor. Thus, pure 3-dimensional Einstein gravity cannot be dual to a LCFT. 

Adding matter fields to Einstein gravity does not help neither. While this may lead to other kinds of LCFTs, it cannot produce a logarithmic partner for the energy momentum tensor. This is so, because the energy momentum tensor corresponds to graviton (spin-2) excitations in the bulk, and the only field producing such excitations is the metric. 

Therefore, what we need is a way to provide additional degrees of freedom in the gravity sector. The most natural way to do this is by considering higher derivative interactions of the metric. The first gravity model of this type was constructed by Deser, Jackiw and Templeton \cite{Deser:1982vy} %,Deser:1982wh,Deser:1982a} 
who introduced a Chern--Simons term for the Christoffel connection. 
\eq{
S_{\textrm{\tiny CS}} = -\frac{1}{16\pi G_N\,\mu}\,\int\volint\,\epsilon^{\la\mu\nu}\Ga^\rho{}_{\si\la}\Big[\partial_\mu\Ga^\si{}_{\rho\nu}+\frac23 \,\Ga^\si{}_{\kappa\mu}\Ga^\kappa{}_{\si\nu}\Big]
}{eq:f3}
Here $\mu$ is a real coupling constant. Adding this action to the Einstein--Hilbert action \eqref{eq:f2} generates massive graviton excitations in the bulk, which is encouraging for our wish list since we need these extra degrees of freedom. The model that arises when summing the actions \eqref{eq:f2} and \eqref{eq:f3}, %$S_{\textrm{\tiny CTMG}}=S_{\textrm{\tiny EH}}+S_{\textrm{\tiny CS}}$, 
\eq{
S_{\textrm{\tiny CTMG}}=S_{\textrm{\tiny EH}}+S_{\textrm{\tiny CS}}
}{eq:angelinajolie}
is known as ``cosmological topologically massive gravity'' (CTMG) \cite{Deser:1982sv}. It was demonstrated by Kraus and Larsen \cite{Kraus:2005zm} that the central charges in CTMG are shifted from their Brown--Henneaux values:
\eq{
c_L=\frac{3\ell}{2G_N}\,\big(1-\frac{1}{\mu\ell}\big)\qquad c_R=\frac{3\ell}{2G_N}\,\big(1+\frac{1}{\mu\ell}\big)
}{eq:LCFT7}
This is again good news concerning our wish list, since $c_L$ can be made vanishing by a (non-singular) tuning of parameters in the action.
\eq{
\mu\ell=1
}{eq:LCFT8}
CTMG \eqref{eq:angelinajolie} with the tuning above \eqref{eq:LCFT8} is known as ``cosmological topologically massive gravity at the chiral point'' (CCTMG). It complies with the first five items on our wish list, but we still have to prove that also the last two wishes are granted. To this end we need to find a suitable partner for the graviton.

\section{Keeping logs in massive gravity}\label{sec:3}

\subsection{Login}

In this section we discuss the evidence for the existence of specific gravity duals to LCFTs that has accumulated over the past two years. We start with the theory introduced above, CCTMG, and we end with a relatively new theory, new massive gravity \cite{Bergshoeff:2009hq}. 

\subsection{Seeds of logs}

Given that we want a partner for the graviton we consider now graviton excitations $\psi$ around the AdS background \eqref{eq:cg20} in CCTMG. 
\eq{
g_{\mu\nu}=g^{\textrm{\tiny AdS}_3}_{\mu\nu}+\psi_{\mu\nu}
}{eq:whatever}
Li, Song and Strominger \cite{Li:2008dq} found a nice way to construct them, and we follow their construction here. Imposing transverse gauge $\nabla_\mu \psi^{\mu\nu}=0$ and defining the mutually commuting first order operators
\eq{
\big({\cal D}^M\big)_\mu^\be = \de_\mu^\be + \frac{1}{\mu}\,\varepsilon_\mu{}^{\al\be}\nabla_\al \qquad \big({\cal D}^{L/R}\big)_\mu^\be = \de_\mu^\be \pm \ell \,\varepsilon_\mu{}^{\al\be}\nabla_\al
}{eq:f22}
allows to write the linearized equations of motion around the AdS background \eqref{eq:cg20} as follows.
\eq{
({\cal D}^M{\cal D}^L{\cal D^R} \psi)_{\mu\nu} = 0
}{eq:LCFT9}
A mode annihilated by ${\cal D}^M$ (${\cal D}^L$) [${\cal D}^R$] \{$({\cal D}^L)^2$ but not by ${\cal D}^L$\} is called massive (left-moving) [right-moving] \{logarithmic\} and is denoted by $\psi^M$ ($\psi^L$) [$\psi^R$] \{$\psi^{\rm log}$\}. Away from the chiral point, $\mu\ell\neq 1$, the general solution to the linearized equations of motion \eqref{eq:LCFT9} is obtained from linearly combining left, right and massive modes \cite{Li:2008dq}. At the chiral point  ${\cal D}^M$ degenerates with ${\cal D}^L$ and the general solution to the linearized equations of motion \eqref{eq:LCFT9} is obtained from linearly combining left, right and logarithmic modes \cite{Grumiller:2008qz}. Interestingly, we discovered in \cite{Grumiller:2008qz} that the modes $\psi^{\rm log}$ and $\psi^L$ behave as follows:
\eq{ 
(L_0+\bar L_0) \left(\begin{array}{c} \psi^{\rm log} \\ \psi^L
\end{array}\right) = \left(\begin{array}{c@{\quad}c}
2 & 1 \\
0 & 2
\end{array}\right) \left(\begin{array}{c} \psi^{\rm log} \\ \psi^L \end{array}\right)
}{eq:cg79a} 
where $L_0=i\partial_u$, $\bar L_0=i\partial_v$ and
\eq{ 
(L_0-\bar L_0) \left(\begin{array}{c} \psi^{\rm log} \\ \psi^L
\end{array}\right) = \left(\begin{array}{c@{\quad}c}
2 & 0 \\
0 & 2
\end{array}\right) \left(\begin{array}{c} \psi^{\rm log} \\ \psi^L \end{array}\right)
}{eq:cg1a}
If we define naturally the Hamiltonian by $H=L_0+\bar L_0$ and the angular momentum by $J=L_0-\bar L_0$ we recover exactly \eqref{eq:cg79} and \eqref{eq:cg1}, which suggests that the CFT dual to CCTMG (if it exists) is logarithmic, as conjectured in \cite{Grumiller:2008qz}. It was further shown with Jackiw that the existence of the logarithmic excitations $\psi^{\rm log}$ is not an artifact of the linearized approach, but persists in the full theory \cite{Grumiller:2008pr}.

Thus, also the sixth wish is granted in CCTMG. The rest of this section discusses the last wish.

\subsection{Growing logs}

We assume now that there is a standard AdS/CFT dictionary \cite{Aharony:1999ti} available for LCFTs and check if CCTMG indeed leads to the correct 2- and 3-point correlators. To this end we have to identify the sources $j_i$ that appear on the right hand side of the correlator equation \eqref{eq:LCFT10}. Following the standard AdS/CFT prescription the sources for the operators ${\cal O}^L$ (${\cal O}^R$) [${\cal O}^{\rm log}$] are given by left (right) [logarithmic] non-normalizable solutions to the linearized equations of motion \eqref{eq:LCFT9}. Thus, our first task is to find all solutions of the linearized equations of motion and to classify them into normalizable and non-normalizable ones, where ``normalizable'' refers to asymptotic (large $\rho$) behavior that is exponentially suppressed as compared to the AdS background \eqref{eq:cg20}. 

A construction of all normalizable left and right solutions was provided in \cite{Li:2008dq}, and the normalizable logarithmic solutions were constructed in \cite{Grumiller:2008qz}.\footnote{All these modes are compatible with asymptotic AdS behavior \cite{Grumiller:2008es,Henneaux:2009pw}, and they appear in vacuum expectation values of 1-point functions. Indeed, the 1-point function $\langle T^{ij}\rangle$ involves both $\psi^{\rm log}$ and $\psi^R$ \cite{Henneaux:2009pw,Maloney:2009ck,Skenderis:2009nt,Ertl:2009ch}.} The non-normalizable solutions were constructed in \cite{Grumiller:2009mw}. It turned out to be convenient to work in momentum space
\eq{
\psi_{\mu\nu}^{L/R/\rm log}(h,\bar h) = e^{-ih(t+\phi)-i\bar h(t-\phi)}\,F_{\mu\nu}^{L/R/\rm log}(\rho)
}{eq:NMG17}
The momenta $h,\bar h$ are called ``weights''. All components of the tensor $F_{\mu\nu}$ are determined algebraically, except for one that is determined from a second order (hypergeometric) differential equation. In general one of the linear combinations of the solutions is singular at the origin $\rho=0$, while the other is regular there. We keep only regular solutions. For each given set of weights $h,\bar h$ the regular solution is either normalizable or non-normalizable. %Normalizable refers to the asymptotic (large $\rho$) behavior and requires that all components of the linearized solution $\psi_{\mu\nu}$ grow at most linearly in $\rho$, but not exponentially. By contrast, in any non-normalizable mode there is at least one component that grows asymptotically like $(a\rho+1)\,e^{2\rho}$, where the constant $a$ vanishes for left and right solutions but is non-vanishing for logarithmic ones.
It turns out that normalizable solutions exist for integer weights $h\geq 2$, $\bar h\geq 0$ (or $h\leq -2$, $\bar h\leq 0$). All other solutions are non-normalizable. 

An example for a normalizable left mode is given by the primary with weights $h=2$, $\bar h=0$
\eq{
\psi^L_{\mu\nu}(2,0) = \frac{e^{-2iu}}{\cosh^4{\!\!\rho}}\left(\begin{array}{c@{\quad}c@{\quad}c}
\frac14\,\sinh^2{\!(2\rho)} & 0 & \frac i2 \sinh{(2\rho)} \\
0 & 0 & 0 \\
\frac i2 \sinh{(2\rho)} & 0 & -1
\end{array}\right)_{\!\!\!\mu\nu} 
}{eq:cg46}
Note that all components of this mode behave asymptotically ($\rho\to\infty$) at most like a constant. The corresponding logarithmic mode is given by
\eq{
\psi^{\rm log}_{\mu\nu}(2,0)= -\frac12\,(i(u+v)+\ln{\cosh^2\!{\rho}})\,\psi^L_{\mu\nu}(2,0)
}{eq:cg26}
Evidently, it behaves asymptotically like its left partner \eqref{eq:cg46}, except for overall linear growth in $\rho$. It is also worthwhile emphasizing that the logarithmic mode \eqref{eq:cg26} depends linearly on time $t=(u+v)/2$. Both features are inherent to all logarithmic modes. All other normalizable modes can be constructed from the primaries \eqref{eq:cg46}, \eqref{eq:cg26} algebraically. 

An example for a non-normalizable left mode is given by the mode with weights $h=1$, $\bar h=-1$
\eq{
\psi^L_{\mu\nu}(1,-1)=\frac14\,e^{-iu+iv}\,\left(\begin{array}{ccc}
0 & 0 & 0 \\
0 & \cosh{(2\rho)}-1 & -2i\sqrt{\frac{\cosh{(2\rho)}-1}{\cosh{(2\rho)}+1}} \\
0 & -2i\sqrt{\frac{\cosh{(2\rho)}-1}{\cosh{(2\rho)}+1}} & -\frac{4}{\cosh{(2\rho)}+1}
\end{array}\right)_{\!\!\mu\nu}
}{eq:Lgrav}
Note that all components of this mode behave asymptotically ($\rho\to\infty$) at most like a constant, except for the $vv$-component, which grows like $e^{2\rho}$. The corresponding logarithmic mode grows again faster than its left partner \eqref{eq:Lgrav} by a factor of $\rho$ and depends again linearly on time.

Given a non-normalizable solution $\psi^L$ obviously also $\alpha\,\psi^L$ is a non-normalizable solution, with some constant $\alpha$. To fix this normalization ambiguity we demand standard coupling of the metric to the stress tensor:
\eq{
S(\psi^{u\,L}_v,T^v_u)=  \frac12\,\int\!\extd t\extd\phi \sqrt{-g^{(0)}}\,\psi^{uu}_{L}T_{uu} = \int\!\extd t\extd\phi\,e^{-ihu-i\bar h v}\,T_{uu}
}{eq:canonicalcoupling}
Here $S$ is either some CFT action with background metric $g^{(0)}$ or a dual gravitational action with boundary metric $g^{(0)}$. The non-normalizable mode $\psi^L$ is the source for the energy-momentum flux component $T_{uu}$. The requirement \eqref{eq:canonicalcoupling} fixes the normalization. The discussion above focussed on left modes. For the right modes essentially the same discussion applies, but with the substitutions $L\leftrightarrow R$, $h\leftrightarrow\bar h$ and $u\leftrightarrow v$.

\subsection{Logging correlators}

Generically the 2-point correlators on the gravity side between two modes
$\psi^1(h,\bar h)$ and $\psi^2(h',\bar h')$ in momentum space are
determined by
\eq{
\langle \psi^1(h,\bar h)\, \psi^2(h',\bar h')\rangle = \frac12
\,\big(\de^{(2)} S_{\textrm{\tiny CCTMG}}(\psi^1,\psi^2)+ \de^{(2)}
S_{\textrm{\tiny CCTMG}}(\psi^2,\psi^1)\big)
}{eq:2pointdef}
where  $\langle\psi^1\,\psi^2\rangle$  stands for the correlation function of the CFT operators dual to the graviton modes $\psi^1$ and $\psi^2$.
On the right hand side one has to plug the non-normalizable modes $\psi^1$
and $\psi^2$ into the second variation of the on-shell action
and symmetrize with respect to the
two modes. The second variation of the on-shell action of CCTMG
\eq{
\de^{(2)} S_{\textrm{\tiny CCTMG}} =
-\frac{1}{16 \pi\, G_N}\,\int\volint\sqrt{-g}\,\big({\cal D}^L\psi^{1\,\ast}\big)^{\mu\nu} \de G_{\mu\nu}(\psi^2) + {\rm boundary\;terms}
}{eq:2p1}
turns out to be very similar to the second variation of the on-shell Einstein--Hilbert action
\eq{
\de^{(2)} S_{\textrm{\tiny EH}} = -\frac{1}{16\pi\,G_N}\,\int\volint\sqrt{-g}\,\psi^{1\,\mu\nu\,\ast} \de G_{\mu\nu}(\psi^2) + {\rm boundary\;terms}
}{eq:2p3}
This similarity allows us to exploit results from Einstein gravity for CCTMG, as we now explain.\footnote{Alternatively, one can follow the program of holographic renormalization, as it was done by Skenderis, Taylor and van Rees \cite{Skenderis:2009nt}. Their results for 2-point correlators agree with the results presented here.}  The bulk term in CCTMG \eqref{eq:2p1} has the same form as in Einstein theory \eqref{eq:2p3} with $\psi^1$ replaced by ${\cal D}^L\psi^1$. Now, consider boundary terms. Possible obstructions to a well-defined Dirichlet boundary value problem can come only from the variation $\de G_{\mu\nu}(\psi^2)$, since ${\cal D}^L$ is a first order operator. Thus any boundary terms appearing  in \eqref{eq:2p1} containing normal derivatives must be identical with those in Einstein gravity upon substituting $\psi^1\to{\cal D}^L\psi^1$. In addition there can be boundary terms which do not contain normal derivatives of the metric. However, it turns out that such terms can at most lead to contact terms in the holographic computation of 2-point functions. The upshot of this discussion is that we can reduce the calculation of all possible 2-point functions in CCTMG to the equivalent calculation in Einstein gravity with suitable source terms. To continue we go on-shell.\footnote{Above by ``on-shell'' we meant that the background metric is AdS$_3$ \eqref{eq:cg20} and therefore a solution of the classical equations of motion. Here by ``on-shell'' we mean additionally that the linearized equations of motion \eqref{eq:LCFT9} hold.}
\eq{
{\cal D}^L\psi^L=0\qquad{\cal D}^L\psi^R=2\psi^R \qquad{\cal D}^L\psi^{\rm log}=-2\psi^L
}{eq:2p2}
These relations together with the comparison between CCTMG \eqref{eq:2p1} and Einstein gravity \eqref{eq:2p3} then establish 
\begin{subequations}
\label{eq:cor2}
\label{eq:correlatorsI}
\begin{align}
&\langle \psi^R(h,\bar h)\, \psi^R(h',\bar h') \rangle_{\textrm{\tiny
CCTMG}} \sim 2 \langle \psi^R(h,\bar h)\, \psi^R(h',\bar h')
\rangle_{\textrm{\tiny EH}} \\
&\langle \psi^L(h,\bar h)\, \psi^L(h',\bar h') \rangle_{\textrm{\tiny
CCTMG}} \sim 0 \\
&\langle \psi^L(h,\bar h)\, \psi^R(h',\bar h') \rangle_{\textrm{\tiny
CCTMG}} \sim 0  \\
&\langle \psi^R(h,\bar h) \,\psi^{\rm log}(h',\bar h')
\rangle_{\textrm{\tiny CCTMG}} \sim 0 \\
&\langle \psi^L(h,\bar h)\, \psi^{\rm log}(h',\bar h')
\rangle_{\textrm{\tiny CCTMG}} \sim -2 \, \langle \psi^L(h,\bar h)
\,\psi^L(h',\bar h') \rangle_{\textrm{\tiny EH}} 
\end{align}
\end{subequations}
Here the sign $\sim$ means equality up to contact terms. Evaluating the right hand sides in Einstein gravity yields
\eq{
\langle \psi^L(h,\bar h)\,\psi^L(h',\bar h') \rangle_{\textrm{\tiny EH}} = \de_{h,h'}\,\de_{\bar h,\bar h'}\,\frac{c_{BH}}{24}\,\frac{h}{\bar h}(h^2-1)\int\limits_{t_0}^{t_1}\extd t
}{eq:EH4}
and similarly for the right modes, with $h\leftrightarrow\bar h$. The quantity $c_{BH}$ is the Brown--Henneaux central charge \eqref{eq:LCFT5}. The calculation of the 2-point correlator between two logarithmic modes cannot be reduced to a correlator known from Einstein gravity. The result is given by \cite{Grumiller:2009mw}
\eq{
\langle \psi^{\rm log}(h,\bar h)\,\psi^{\rm log}(h',\bar h') \rangle_{\textrm{\tiny CCTMG}} \sim
-\de_{h,h'}\,\de_{\bar h,\bar h'}\,\frac{\ell}{4\,G_N}\,\frac{h}{\bar h}\,(h^2-1)\, \big( \psi(h-1) + \psi(-\bar h) \big)\,\int\limits_{t_0}^{t_1}\extd t
}{eq:loglog4}
where $\psi$ is the digamma function. An ambiguity in defining $\psi^{\rm log}$, viz., $\psi^{\rm log}\to\psi^{\rm log}+\ga\,\psi^L$, was fixed conveniently in the result \eqref{eq:loglog4}. This ambiguity corresponds precisely to the ambiguity of the LCFT mass scale $m_L$ in \eqref{eq:loglog} (see also the discussion below that equation).

To compare the results \eqref{eq:cor2}-\eqref{eq:loglog4} with the Euclidean 2-point correlators in the short-distance limit \eqref{eq:LCFT1}, \eqref{eq:2point} we take the limit of large weights $h,-\bar h\to\infty$ (e.g.~$\lim_{h\to\infty}\psi(h)=\ln h +{\cal O}(1/h)$) and Fourier-transform back to coordinate space (e.g.~$h^3/\bar h$ is Fourier-transformed into $\partial_z^4/(\partial_z\partial_{\bar z})\,\de^{(2)}(z,\bar z)\propto\partial_z^4\,\ln{|z|}\propto 1/z^4$). Straightforward calculation establishes perfect agreement with the LCFT correlators \eqref{eq:LCFT1}, \eqref{eq:2point}, provided we use the values
\eq{
c_L=0\qquad\qquad c_R=\frac{3\ell}{G_N}\qquad\qquad b_L = -\frac{3\ell}{G_N}
}{eq:cb}
These are exactly the values for central charges $c_L$, $c_R$ \cite{Kraus:2005zm} and new anomaly $b_L$ \cite{Skenderis:2009nt,Grumiller:2009mw} found before. Thus, at the level of 2-point correlators CCTMG is indeed a gravity dual for a LCFT.

\begin{figure}
\begin{center}
\epsfig{file=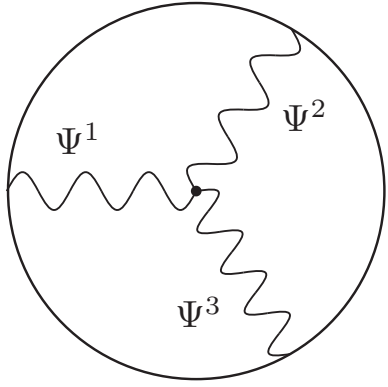,width=0.4\linewidth}
\caption{Witten diagram for three graviton correlator}
\label{fig:1} 
\end{center}
\end{figure}
We evaluate now the Witten diagram in Fig.~\ref{fig:1}, which yields the 3-point correlator
 on the gravity side between three modes $\psi^1(h,\bar h)$, $\psi^2(h',\bar h')$ and $\psi^3(h'',\bar h'')$ in momentum space.
\eq{
\langle \psi^1(h,\bar h)\, \psi^2(h',\bar h')\, \psi^3(h'',\bar h'')\rangle = \frac16 \,\big(\de^{(3)} S_{\textrm{\tiny CCTMG}}(\psi^1,\psi^2,\psi^3)+5\;{\rm permutations}\big)
}{eq:3pointdef}
On the right hand side one has to plug the non-normalizable modes $\psi^1$, $\psi^2$ and $\psi^3$ into the third variation of the on-shell action and symmetrize with respect to all three modes. 
\eq{
\de^{(3)} S_{\textrm{\tiny CCTMG}} \sim -\frac{1}{16\pi\,G_N}\,\int\volint\sqrt{-g}\,\Big[\big({\cal D}^L\psi^1\big)^{\mu\nu}\,\de^{(2)} R_{\mu\nu}(\psi^2,\psi^3) + \psi^{1\,\mu\nu}\,\De_{\mu\nu}(\psi^2,\psi^3)\Big]
}{eq:3p3}
The quantity $\de^{(2)} R_{\mu\nu}(\psi^2,\psi^3)$ denotes the second variation of the Ricci-tensor and the tensor $\De_{\mu\nu}(\psi^2,\psi^3)$ vanishes if evaluated on left- and/or right-moving solutions. All boundary terms turn out to be contact terms, which is why only bulk terms are present in the result \eqref{eq:3p3} for the third variation of the on-shell action. We compare again with Einstein gravity.
\eq{
\de^{(3)} S_{\textrm{\tiny EH}} \sim -\frac{1}{16\pi\,G_N}\,\int\volint\sqrt{-g}\,\psi^{1\,\mu\nu}\,\de^{(2)} R_{\mu\nu}(\psi^2,\psi^3)
}{eq:aha}
Once more we can exploit some results from Einstein gravity for CCTMG, and we find the following results \cite{Grumiller:2009mw} for 3-point correlators without log-insertions:
\begin{subequations}
\label{eq:3cor0log}
\begin{align}
%\!\!\!\!
&\langle \psi^R(h,\bar h)\, \psi^R(h',\bar h')\,\psi^R(h'',\bar h'') \rangle_{\textrm{\tiny CCTMG}} \sim 2\, \langle \psi^R(h,\bar h)\, \psi^R(h',\bar h')\,\psi^R(h'',\bar h'') \rangle_{\textrm{\tiny EH}} \\
%\!\!\!\!
&\langle \psi^L(h,\bar h)\, \psi^R(h',\bar h')\,\psi^R(h'',\bar h'') \rangle_{\textrm{\tiny CCTMG}}  \sim 0 \\
%\!\!\!\!
&\langle \psi^L(h,\bar h)\, \psi^L(h',\bar h')\,\psi^R(h'',\bar h'') \rangle_{\textrm{\tiny CCTMG}} \sim 0 \\
%\!\!\!\!
&\langle \psi^L(h,\bar h)\, \psi^L(h',\bar h')\,\psi^L(h'',\bar h'') \rangle_{\textrm{\tiny CCTMG}} \sim 0
\end{align}
\end{subequations}
with one log-insertion:
\begin{subequations}
\label{eq:3cor1log}
\begin{align}&\!\!\langle \psi^R(h,\bar h)\, \psi^R(h',\bar h')\,\psi^{\rm log}(h'',\bar h'') \rangle_{\textrm{\tiny CCTMG}} \sim 0 \\
&\!\!\langle \psi^L(h,\bar h)\, \psi^R(h',\bar h')\,\psi^{\rm log}(h'',\bar h'') \rangle_{\textrm{\tiny CCTMG}} \sim 0 \\
&\!\!\langle \psi^L(h,\bar h)\, \psi^L(h',\bar h')\,\psi^{\rm log}(h'',\bar h'') \rangle_{\textrm{\tiny CCTMG}} \sim -2\, \langle \psi^L(h,\bar h)\, \psi^L(h',\bar h')\,\psi^L(h'',\bar h'') \rangle_{\textrm{\tiny EH}} 
\end{align}
\end{subequations}
and with two or more log-insertions:
\begin{subequations}
\label{eq:3cor2log}
\begin{align}
&\lim_{|\textrm{weights}|\to\infty}\langle \psi^R(h,\bar h)\, \psi^{\rm log}(h',\bar h')\,\psi^{\rm log}(h'',\bar h'') \rangle_{\textrm{\tiny CCTMG}} \sim 0  \\
&\lim_{|\textrm{weights}|\to\infty}\langle \psi^L(h,\bar h)\, \psi^{\rm log}(h',\bar h')\,\psi^{\rm log}(h'',\bar h'') \rangle_{\textrm{\tiny CCTMG}} \sim \de_{h'',-h-h'}\de_{\bar h'',-\bar h-\bar h'}\,\frac{P^{\rm log}(h,h^\prime,\bar h,\bar h^\prime)}{\bar h \bar h^\prime(\bar h+\bar h^\prime)} \\
&\lim_{|\textrm{weights}|\to\infty}\langle \psi^{\rm log}(h,\bar h)\, \psi^{\rm log}(h',\bar h')\,\psi^{\rm log}(h'',\bar h'') \rangle_{\textrm{\tiny CCTMG}} \sim \de_{h'',-h-h'}\de_{\bar h'',-\bar h-\bar h'}\,\frac{{\rm lengthy}}{\bar h \bar h^\prime(\bar h+\bar h^\prime)} \label{eq:logloglog}
\end{align}
\end{subequations}
The last two correlators so far could be calculated qualitatively only ($P^{\rm log}$ is a known polynomial in the weights and also contains logarithms in the weights, as expected on general grounds), and it would be interesting to calculate them exactly. They are in qualitative agreement with corresponding LCFT correlators. All other correlators have been calculated exactly \cite{Grumiller:2009mw}, and they are in precise agreement with the LCFT correlators \eqref{eq:LCFT1}, \eqref{eq:3point}, provided we use again the values \eqref{eq:cb} for central charges and new anomaly.

In conclusion, also the seventh wish is granted for CCTMG.\footnote{The sole caveat is that two of the ten 3-point correlators were calculated only qualitatively. It would be particularly interesting to calculate the correlator between three logarithmic modes \eqref{eq:logloglog}, since it contains an additional parameter independent from the central charges and new anomaly that determines LCFT properties.} Thus, there are excellent chances that CCTMG is dual to a LCFT with values for central charges and new anomaly given by \eqref{eq:cb}.

\subsection{Logs don't grow on trees}

From the discussion above it is clear that possible gravity duals for LCFTs are sparse in theory space: Einstein gravity \eqref{eq:f2} does not provide a gravity dual for any tuning of parameters and CTMG \eqref{eq:angelinajolie} does potentially provide a gravity dual only for a specific tuning of parameters \eqref{eq:LCFT8}. Any candidate for a novel gravity dual to a LCFT is therefore welcomed as a rare entity.

Very recently another plausible candidate for such a gravitational theory was found \cite{Grumiller:2009sn}. That theory is known as ``new massive gravity'' \cite{Bergshoeff:2009hq}. 
\eq{
S_{\textrm{\tiny NMG}}=\frac{1}{16\pi\,G_N}\,\int\extd^3x\sqrt{-g}\,\Big[\si R+\frac{1}{m^2}\,\big(R^{\mu\nu}R_{\mu\nu}-\frac38\,R^2\big)-2\la m^2\Big]
}{eq:NMG1} 
Here $m$ is a mass parameter, $\lambda$ a dimensionless cosmological parameter and $\sigma=\pm 1$ the sign of the Einstein-Hilbert term. If they are tuned as follows
\eq{
\la \ = \ 3 \qquad\Rightarrow\qquad m^2 \ = \ -\frac{\si}{2\ell^2} 
}{eq:NMG4}
then essentially the same story unfolds as for CTMG at the chiral point. The main difference to CCTMG is that both central charges vanish in new massive gravity at the chiral point (CNMG) \cite{Liu:2009bk,Bergshoeff:2009aq}.
\eq{
c_L =  c_R =\frac{3\ell}{2G_N}\,\left(\sigma+\frac{1}{2\ell^2m^2}\right) = 0
}{eq:NMG6}
Therefore, both left and right flux component of the energy momentum tensor acquire a logarithmic partner. It is easy to check that CNMG grants us the first six wishes from section \ref{sec:2}. The seventh wish requires again the calculation of correlators. The 3-point correlators have not been calculated so far, but at the level of 2-point correlators again perfect agreement with a LCFT was found, provided we use the values \cite{Grumiller:2009sn}
\eq{
c_L=c_R=0\qquad b_L=b_R=-\sigma\,\frac{12\ell}{G_N}
}{eq:bbc}

It is likely that a similar story can be repeated for general massive gravity \cite{Bergshoeff:2009hq}, which combines new massive gravity \eqref{eq:NMG1} with a gravitational Chern--Simons term \eqref{eq:f3}. Thus, even though they are sparse in theory space we have found a few good candidates for gravity duals to LCFTs: cosmological topologically massive gravity, new massive gravity and general massive gravity. In all cases we have to tune parameters in such a way that a ``chiral point'' emerges where at least one of the central charges vanishes.

\subsection{Chopping logs?}

So far we were exclusively concerned with finding gravitational theories where logarithmic modes can arise. In this subsection we try to get rid of them. The rationale behind the desire to eliminate the logarithmic modes is unitarity of quantum gravity. Gravity in 2+1 dimensions is simple and yet relevant, as it contains black holes \cite{Banados:1992wn}, possibly gravity waves \cite{Deser:1982vy} and solutions that are asymptotically AdS. Thus, it could provide an excellent arena to study quantum gravity in depth provided one is able to come up with a consistent (unitary) theory of quantum gravity, for instance by constructing its dual (unitary) CFT. Indeed, two years ago Witten suggested a specific CFT dual to 3-dimensional quantum gravity in AdS \cite{Witten:2007kt}. This proposal engendered a lot of further research (see \cite{Manschot:2007zb,Gaiotto:2007xh,Gaberdiel:2007ve,Yin:2007gv,Yin:2007at,Maloney:2007ud,Manschot:2007ha} for some early references), including the suggestion by Li, Song and Strominger \cite{Li:2008dq} to construct a quantum theory of gravity that is purely right-moving, dubbed ``chiral gravity''. To make a long story \cite{Carlip:2008jk,Grumiller:2008pr,Carlip:2008qh,Grumiller:2008qz,Hotta:2008yq,Li:2008yz,Park:2008yy,Sachs:2008gt,Lowe:2008ye,Myung:2008ey,Carlip:2008eq,Lee:2008gta,Sachs:2008yi,Gibbons:2008vi,Anninos:2008fx,Giribet:2008bw,Strominger:2008dp,Compere:2008cv,Myung:2008dm,deHaro:2008gp,Stevens:2008hv,Deser:2008rm,Hotta:2008xt,Quevedo:2008ry,Oh:2008tc,Garbarz:2008qn,Kim:2008bf,Mann:2008rx,Blagojevic:2008bn,Nam:2009dd,Hellerman:2009bu,Sezgin:2009dj,Anninos:2009zi,Compere:2009zj,Hotta:2009zn,Anninos:2009jt,Carlip:2009ey,Chow:2009km,Becker:2009mk,Blagojevic:2009ek,Vasquez:2009mk,Duncan:2009sq,Andrade:2009ae,Miskovic:2009kr,Skenderis:2009kd,Ertl:2009ch,Afshar:2009rg} short, ``chiral gravity'' is nothing but CCTMG with the logarithmic modes truncated in some consistent way.

We discuss now two conceptually different possibilities of implementing such a truncation. The first option was proposed in \cite{Grumiller:2008qz}. If one imposes periodicity in time for all modes, $t\to t+\beta$, then only the left- and right-moving modes are allowed, while the logarithmic modes are eliminated since they grow linearly in time, see e.g.~\eqref{eq:cg26}. %Such boundary conditions could be considered as natural for finite temperature applications, so we may paraphrase the first option as ``logs on fire may evaporate''. 
The other possibility was pursued in \cite{Maloney:2009ck}. It is based upon the observation that logarithmic modes grow logarithmically faster in $e^{2\rho}$ than their left partners, see e.g.~\eqref{eq:cg26}. Thus, imposing boundary conditions that prohibit this logarithmic growth eliminates all logarithmic modes. %In neither of these cases is it clear why the truncation should remain consistent in the full quantum theory, so the jury on 3-dimensional quantum gravity is still out.

Currently it is not known whether chiral gravity has its own dual CFT or if it exists merely as a zero-charge superselection sector of the logarithmic CFT. In the latter case it is unclear whether or not the zero-charge superselection sector is a fully-fledged CFT. Another alternative is that neither the LCFT nor its chiral truncation dual to chiral gravity exists. In that case CTMG is unlikely to exist as a consistent quantum theory on its own. Rather, it would require a UV completion, such as string theory.

\subsection{Logout}

We summarize now the key results reviewed in this section as well as some open issues. Cosmological topologically massive gravity \eqref{eq:angelinajolie} at the chiral point \eqref{eq:LCFT8} is likely to be dual to a LCFT with a logarithmic partner for one flux component of the energy momentum tensor since 2- \cite{Skenderis:2009nt} and 3-point correlators \cite{Grumiller:2009mw} match. The values of central charges and new anomaly are given by \eqref{eq:cb}. The detailed calculation of the correlator with three log-insertions \eqref{eq:logloglog} still needs to be performed and will determine another parameter of the LCFT. New massive gravity \eqref{eq:NMG1} at the chiral point \eqref{eq:NMG4}  is likely to be dual to a LCFT with a logarithmic partner for both flux components of the energy momentum tensor since 2-point correlators match \cite{Grumiller:2009sn}.  The central charges vanish and the new anomalies are given by \eqref{eq:bbc}. The calculation of 3-point correlators still needs to be performed and will provide a more stringent test of the conjectured duality to a LCFT. A similar story is likely to repeat for general massive gravity (the combination of topologically and new massive gravity) at a chiral point, and it could be rewarding to investigate this issue. Finally we addressed possibilities to eliminate the logarithmic modes and their partners, since such an elimination might lead to a chiral theory of quantum gravity \cite{Li:2008dq}, called ``chiral gravity''. The issue of whether chiral gravity exists still remains open.

\section{Towards condensed matter applications}\label{sec:4}

In this final section we review briefly some condensed matter systems where LCFTs do arise, see \cite{Flohr:2001zs,Gaberdiel:2001tr} for more comprehensive reviews. We focus on LCFTs where the energy-momentum tensor acquires a logarithmic partner, i.e., the class of LCFTs for which we have found possible gravity duals.\footnote{A well-studied alternative case is a LCFT with $c=-2$ \cite{Gurarie:1993xq,Gaberdiel:1996np}. There is no obvious way to construct a gravity dual for such LCFTs, even when considering CTMG or new massive gravity away from the chiral point. We thank Ivo Sachs for discussions on this issue.} Condensed matter systems described by such LCFTs are for instance systems at (or near) a critical point with quenched disorder, like spin glasses \cite{Binder:1986zz}/quenched random magnets \cite{Cardy:1999zp,RezaRahimiTabar:2000qr}, dilute self-avoiding polymers or percolation \cite{Gurarie:1999yx}. ``Quenched disorder'' arises in a condensed matter system with random variables that do not evolve with time. If the amount of disorder is sufficiently large one cannot study the effects of disorder by perturbing around a critical point without disorder --- standard mean field methods break down. The system is then driven towards a random critical point, and it is a challenge to understand its precise nature. Mathematically, the essence of the problem lies in the infamous denominator arising in correlation functions of some operator ${\cal O}$ averaged over disordered configurations (see e.g.~chapter VI.7 in \cite{Zee:2003mt})
\eq{
\overline{\langle{\cal O}(z)\,{\cal O}(0)\rangle} = \int {\cal D} V P[V]\, \frac{\int{\cal D}\phi\,\exp{\big(-S[\phi]-\int\extd^2 z' V(z'){\cal O}(z')\big)}\,{\cal O}(z)\,{\cal O}(0)}{\int{\cal D}\phi\,\exp{\big(-S[\phi]-\int\extd^2 z' V(z'){\cal O}(z')\big)}}
}{eq:quencor}
Here $S[\phi]$ is some 2-dimensional\footnote{Analog constructions work in higher dimensions, but we focus here on two dimensions.} quantum field theory action for some field(s) $\phi$ and $V(z)$ is a random potential with some probability distribution. For white noise one takes the Gaussian probability distribution $P[V]\propto \exp{\big(-\int d^2z V^2(z)/(2g^2)\big)}$, where $g$ is a coupling constant that measures the strength of the impurities. If it were not for the denominator appearing on the right hand side of the averaged correlator \eqref{eq:quencor} we could simply perform the Gaussian integral over the impurities encoded in the random potential $V(z)$. This denominator is therefore the source of all complications and to deal with it requires suitable methods, see e.g.~\cite{Bernard:1995as}. One possibility is to eliminate the denominator by introducing ghosts. This so-called ``supersymmetric method'' works well if the original quantum field theory described by the action $S[\phi]$ is very simple, like a free field theory. Another option is the so-called replica trick, where one introduces $n$ copies of the original quantum field theory, calculates correlators in this setup and takes the limit $n\to 0$ in the end, which formally reproduces the denominator in \eqref{eq:quencor}. Recently, Fujita, Hikida, Ryu and Takayanagi combined the replica method with the AdS/CFT correspondence to describe disordered systems \cite{Fujita:2008rs} (see \cite{Kiritsis:2008at,Myers:2008me} for related work), essentially by taking $n$ copies of the CFT, exploiting AdS/CFT to calculate correlators and taking formally the limit $n\to 0$ in the end. Like other replica tricks their approach relies on the existence of the limit $n\to 0$.

One of the results obtained by the supersymmetric method or replica trick
is that correlators like the one in \eqref{eq:quencor} develop a logarithmic behavior, exactly as in a LCFT \cite{Cardy:1999zp}. In fact, in the $n\to 0$ limit prescribed by the replica trick, the conformal dimensions of certain operators degenerate. This produces a Jordan block structure for the Hamiltonian in precise parallel to the $\mu\ell \to 1$ limit of CTMG. More concretely, LCFTs can be used to compute correlators of quenched random systems!

This suggests yet-another route to describe systems with quenched disorder,
and our present results add to this toolbox.
Namely, instead of taking $n$ copies of an ordinary
CFT we may start directly with a LCFT. If this LCFT is weakly
coupled we can work on the LCFT side perturbatively, using the results
mentioned above \cite{Cardy:1999zp,Gurarie:1999yx,RezaRahimiTabar:2000qr,Flohr:2001zs,Gaberdiel:2001tr}.
On the other hand, if the LCFT becomes strongly coupled, perturbative
methods fail. To get a handle on these situations we can exploit the
AdS/LCFT correspondence and work on the gravity side. Of course, to this
end one needs to construct gravity duals for LCFTs. The models reviewed in
this talk are simple and natural examples of such constructions.

%Our results suggest yet-another route to describe systems with quenched disorder. Namely, instead of taking $n$ copies of an ordinary CFT we may start directly with a LCFT. If this LCFT is weakly coupled we can work on the LCFT side perturbatively, using the results mentioned above \cite{Cardy:1999zp,Gurarie:1999yx,RezaRahimiTabar:2000qr,Flohr:2001zs,Gaberdiel:2001tr}. If the LCFT becomes strongly coupled we exploit the AdS/LCFT correspondence and work on the gravity side. To this end one needs to construct gravity duals for LCFTs, perhaps similar to the ones reviewed in this talk.

%\newpage

\ack

We thank Matthias Gaberdiel, Gaston Giribet, Olaf Hohm, Roman Jackiw, David Lowe, Hong Liu, Alex Maloney, John McGreevy, Ivo Sachs, Kostas Skenderis, Wei Song, Andy Strominger and Marika Taylor for discussions.
DG thanks the organizers of the ``First Mediterranean Conference on Classical and Quantum Gravity'' for the kind invitation and for all their efforts to make the meeting very enjoyable.
DG and NJ are supported by the START project Y435-N16 of the Austrian Science Foundation (FWF). During the final stage NJ has been supported by project P21927-N16 of FWF. NJ acknowledges financial support from the Erwin-Schr\"odinger-Institute (ESI) during the workshop ``Gravity in three dimensions''.

%\newpage

\section*{References}

%\bibliographystyle{iopart-num}
%\bibliography{review} 

\providecommand{\newblock}{}

\end{document}